# Non-Transferability in Communication Channels and Tarski's Truth Theorem

Farhad Naderian

System Integration Specialist, Freelance; Tehran, Iran

naderian51@gmail.com

ORCID:0000-0002-1771-2325

**Abstract**

This article explores the concept of transferability within communication channels, with a particular focus on the inability to transmit certain situations through these channels. The Channel Non-Transferability Theorem establishes that no encoding-decoding mechanism can fully transmit all propositions, along with their truth values, from a transmitter to a receiver. The theorem underscores that when a communication channel attempts to transmit its own error state, it inevitably enters a non-transferable condition. I argue that Tarski's Truth Undefinability Theorem parallels the concept of non-transferability in communication channels. As demonstrated in this article, the existence of non-transferable codes in communication theory is mathematically equivalent to the undefinability of truth as articulated in Tarski's theorem. This equivalence is analogous to the relationship between the existence of non-computable functions in computer science and Gödel's First Incompleteness Theorem in mathematical logic. This new perspective sheds light on additional aspects of Tarski's theorem, enabling a clearer expression and understanding of its implications.

**Keywords:** Non-Transferability, Channel Theory, Tarski's Truth Theorem, Semantic.

# 1. Introduction

In modern telecommunications, data transmission relies on the standard model of communication proposed by Shannon and Weaver [1]. As shown in *Figure 1* , this traditional model comprises three main components.

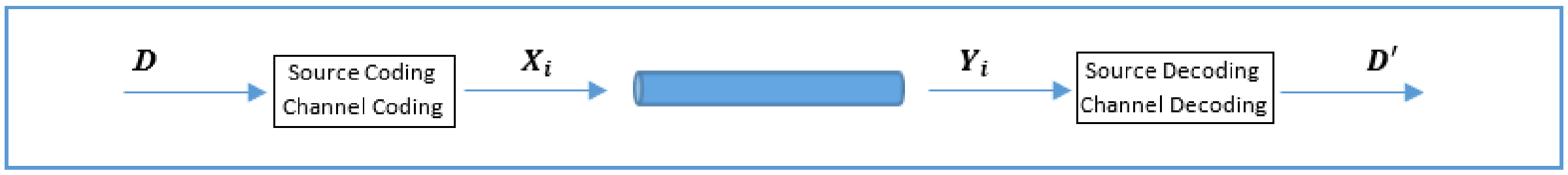

**Fig.1 The traditional scheme of a communication channel**

1. ***Source/Channel Coding**:* It is responsible for getting data **D** and converting it to the suitable string $\boldsymbol{X_i}$.
2. ***Channel:*** It takes String $\boldsymbol{X_i}$ and sends it toward the receiver.
3. ***Source/Channel Decoding:*** It constructs the final data $\boldsymbol{D'}$ based on the received signal $\boldsymbol{Y_i}$ from the channel.

In an ideal communication channel, $\boldsymbol{D' = D}$**,** meaning the final received data is identical or as close as possible to the transmitted data [2].

This model primarily addresses syntax, not the origin or meaning of data $\boldsymbol{D}$. Consequently, it cannot address semantic issues unless the source of $\boldsymbol{D}$ is defined. To account for meaning, the model must be expanded appropriately.

This article aims to address semantic issues by introducing a new concept: "non-transferability of meaning". This concept, previously unaddressed, is introduced here for the first time[1]. Additionally, the article explores a fundamental question about the "existence of non-transferable situations in communication channels."

Investigating the non-transferability of meaning in communication channels depends on foundational research into the semantics and logic of meaning. Fortunately, the concept of meaning has been extensively examined from both logical and semantic perspectives [3]. The article builds on these studies to examine non-transferability issues in communication channels and establishes its relationship with a significant result in mathematical logic.

One of the key restrictive results in mathematical logic is Tarski's undefinability theorem, which states that there is no formal method to define "Truth" within the language of arithmetic [4]. Initially formulated to address the semantic concept of truth in mathematical logic, this theorem has not been explored in other fields until now. This article draws a parallel between Tarski's theorem and non-transferability in communication channels, finding a similarity to the relationship between Gödel's incompleteness theorem and halting problem in Turing machines. Gödel's incompleteness theorem reveals that some true propositions in arithmetic cannot be proven within arithmetic itself, analogous to some codes that, when given to a universal Turing machine, cause it not to halt. This relationship connects mathematical logic and computer science and clarified by Scott Aaronson in his 2011 blog at **"Rosser's theorem via Turing machines. Shtetl-Optimized"**. Likewise, Tarski's truth theorem can be seen as connected to non-transferability issues in communication channels, suggesting that the undefinability of truth in logic aligns with limitations in communication theory.

[1] The term "non-transferability" is used in this article in a new meaning for the first time. A formal definition of it is at the end of Section 2.

In the following, Section 2 will provide foundational definitions to establish the groundwork for the article, including a generalization of the communication model to address the semantic issues. A definition of transferability in communication channels appear at the end of Section 2. The occurrence of non-transferable circumstances in communication channels is demonstrated in Section 3. The article's central claim—that non-transferability in communication channels is equivalent to the undefinability of truth in arithmetic—is presented in Section 4. The translation of the Liar Paradox into communication theory, an immediate corollary of this theorem, is also covered in this section. The findings and conclusions of this study are summed up in Section 5.

## 2. Communication channel model

In practical communication channels, data originates from real-world scenarios or sets of facts. For instance, it may represent the switching status of devices in the utility system, alarms within a telecommunication network, or measurements collected by sensor arrays. Additionally, it could involve signals derived from real-time sources such as voice, images or video. While some data sources are inherently analog, they are converted into digital streams to facilitate transmission over digital channels. In *Figure 2* , a communication channel scheme is represented to model semantic issues. In this figure, set or situation **S** refers to some state of affairs in the world. The communication channel must be able to send those states of affairs to the receiver.

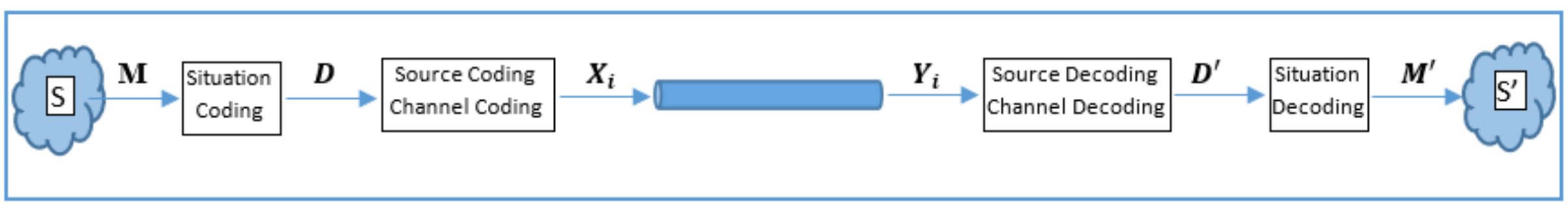

**Fig.2 A communication channel scheme to model semantic issues**

There are various methods to represent these states of affairs; however, for the purposes of this paper, they will be considered as sets of simple first-order sentences expressed through unary (**1-ary)** relations or properties assigned to different objects denoted as $\#\boldsymbol{m}$:

$$\begin{aligned} &M = \{P_1(1), \neg P_1(1), P_1(2), \neg P_1(2), \ldots, P_n(m), \neg P_n(m), \ldots\} \\ &P_n(m): \quad \textit{Object \#m has property } P_n \\ &\neg P_n(m): \quad \textit{Object \#m has not property } P_n \end{aligned} \tag{2.1}$$

Examples of simple properties are Device-OK($\mathbf{1}$), Battery-Low($\mathbf{4}$), AC-Fail($\mathbf{14}$), Switch-On($\mathbf{145}$) which represent the operational state of device #$\mathbf{1}$, Low battery alarm for device #$\mathbf{4}$, a failure in the AC system of device #$\mathbf{14}$, or the **On** State of Switch #$\mathbf{145}$, respectively**.** While representing more complex facts such as those found in extensive databases, requires **n-ary** relations and advanced logical structures, including quantifiers and model operators, the use of simple properties and their negations, as shown above, suffices for the scope of this paper

To transmit the informational content of a situation $\boldsymbol{S}$, it is modeled by a set of **1-ary** properties as in 2.1. Each constituent proposition is then represented by a frame, as will be described in Definition 2.1 below, and these frames are sequentially transmitted to the receiver. In this article, each proposition within the message is encoded using a binary frame. An appropriate alphabet is selected to represent all symbols, which are then converted into binary strings using binary codes, such as ASCII. For numerical data, the binary representation of the numbers is utilized. Below, I define the encoding procedure for a **1-ary** predicate.

***Definition 2.1***: The proposition "$\boldsymbol{P_n(m) = Object\ \#m\ has\ property\ P_n}$" is encoded with the data frame $\boldsymbol{D = F_{P_n(m)}}$ by the following rules based on *Figure 3* :

a) $\boldsymbol{D = F_{P_n(m)}}$ is the binary code of $\boldsymbol{P_n(m)}$**.** That is the concatenation of strings: **Polarity**, $\boldsymbol{Code(P_n)}$, and $\boldsymbol{Code(m)}$**.**
b) $\boldsymbol{Code(m)}$ is the binary representation of the number $\boldsymbol{m}$ in order to code $\boldsymbol{Object\ \#m}$**.**
c) $\boldsymbol{Code(P_n)}$ is the binary representation of the ASCII code for $\boldsymbol{P_n}$ (or the number $\boldsymbol{n}$**)** in order to code predicate**.**
d) **Polarity** is the truth value of $\boldsymbol{P_n(m)}$. It is **1** if $\boldsymbol{P_n(m)}$ is true and is **0** if $\boldsymbol{P_n(m)}$ is false.
e) $\boldsymbol{x_0}$ or **0** is to refer to all possible objects $\boldsymbol{m}$.
f) $\boldsymbol{P_n(0)}$ is the proposition to represent "all objects have property $\boldsymbol{P_n}$".
g) Another simple representation of the frame is the triple $\boldsymbol{D}$=**[Polarity,** $\boldsymbol{Code(P_n)}$, $\boldsymbol{Code(m)}$**].■**

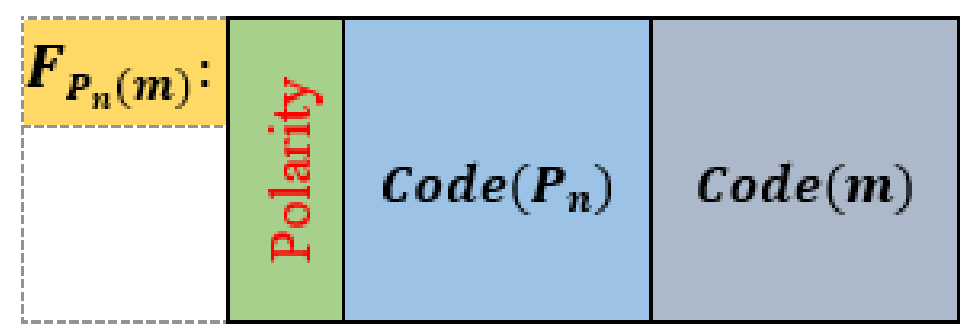

**Fig.3 Each proposition$P_n(m)$ can be represented by a string that is the concatenation of three strings for Polarity, the code for $P_n$, and the code for m, or as the triple[Polarity, $Code(P_n)$, $Code(m)$]**

To ensure unambiguous decoding, the codes of $\boldsymbol{m}$ and $\boldsymbol{P_n}$ must be unique. This can be achieved by taking ASCII symbols to represent properties or by employing binary representations of objects. Additionally, a unique string ($\boldsymbol{x_0}$ or **0)** is necessary to reference to all values in predicates, a practice common in networking and communication protocols[2]. It is important to note that the $\boldsymbol{F_{P_n(m)}}$ itself forms a string, which can be generated if a standardized scanning method, such as left to right, is adopted. In practical communication and networking setups, frames often include additional components to convey supplementary information about the data, aiding tasks like frame alignment, error detection and correction, data type specification, and frame length management.

As an example of a frame, let's encode the proposition **ON(112)** to show the fact that **System#112** is **active**. The ASCII code of the property **ON** is 4F|4E = 01001111|01001110. The code of object **#112** is 1110000. Then $\boldsymbol{F_{ON(112)} = 1|0100111101001110|1110000|}$ or the triple (1,4F4E,112) as in *Figure 4 :*

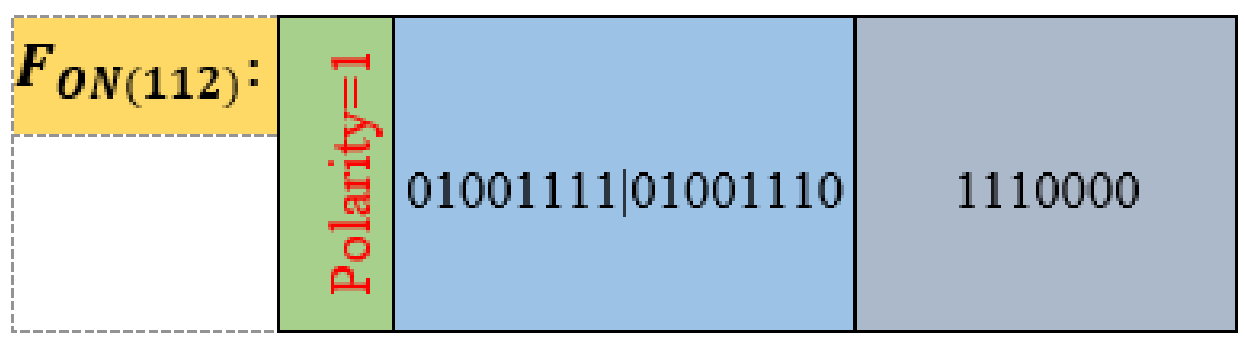

**Fig.4 Frame representation of the proposition ON(112)**

For the purpose of this article, it is essential to ensure that the Situation Coder-Decoder has mechanisms in place to unambiguously distinguish the three parts of the frame[3].

---

[2] In many networking protocols, to refer to all quantities of a variable, all-0 or all-1 is used. For example, the IP address with all 1 or 255.255.255.255 is used to refer to all IP addresses.

[3] The coding methods that has described, is based on a simple model of data frames. There are many techniques in networking and telecommunication to reach an unambiguous coding-decoding and I prevent referring to them.

***Definition 2.2***: An active Communication Channel $\boldsymbol{C}$ is a triple $(\boldsymbol{Encode}(.), \boldsymbol{TS}(.), \boldsymbol{Decode}(.))$ in which:

- $\boldsymbol{Encode}(.)$ is a function from the set of **1-ary** predicates $\boldsymbol{P_n}(\boldsymbol{m})$ into the data streams $\boldsymbol{D} = \boldsymbol{F_{P_n(m)}}$ (based on definition 2.1)
- $\boldsymbol{TS}(.)$ or **T**ransmission **S**ystem is the function that takes data String $\boldsymbol{D} = \boldsymbol{F_{P_n(m)}}$ and deliver it to receiver as $\boldsymbol{D'}$.
- $\boldsymbol{Decode}(.)$ is the function that takes received data $\boldsymbol{D'}$ and builds a proposition $\boldsymbol{P'_{n'}}(\boldsymbol{m'})$ according to rules in definition 2.1 .
- $\boldsymbol{TS}(.)$ can transmit symbols of data. It means that both channel coding and source coding can be selected in such a way that a one-to-one relationship between $\boldsymbol{X_i}$ and $\boldsymbol{Y_i}$ is possible. This condition ensures the activeness of the channel. $\boldsymbol{TS}(.)$ , in fact, includes both source coding and channel coding.

***Definition 2.3***: A proposition is called *transferable* over a channel if it is logically equivalent to the received proposition from the channel on which it has been transmitted. If the received proposition is not equivalent to the transmitted one, then that proposition is *non-transferable* over that channel:

Let's channel $\boldsymbol{C}$ is $(\boldsymbol{Encode}(.), \boldsymbol{TS}(.), \boldsymbol{Decode}(.))$
Then a proposition $\boldsymbol{P_n}(\boldsymbol{m})$ is *transferrable* over the channel $\boldsymbol{C}$ if and only if: (2.2)

$$\boldsymbol{Decode}\left(\boldsymbol{TS}\left(\boldsymbol{Encode}\left(\boldsymbol{P_n}(\boldsymbol{m})\right)\right)\right) \leftrightarrow \boldsymbol{P_n}(\boldsymbol{m}) \;\blacksquare$$

## 3. Non-Transferability Theorem in Communication Channels

With the necessary Coding-Decoding mechanisms in communication channels defined we are now prepared to present the theorem related to transferability within these channels. This section includes both an informal version of the move and a formula one.

***Theorem 3.1****:* An active communication channel cannot encode and send every proposition toward the receiver correctly, or not all propositions are transferable over an active communication channel.

***Proof A, (****Informal setup):*

Based on the structure of the introduced communication channel, every proposition in the system can be encoded and transmitted on the channel. Let's construct a proposition named $\boldsymbol{NT}(\boldsymbol{0})$ and send it toward the receiver as in *Figure 5* . The meaning of $\boldsymbol{NT}(\boldsymbol{x})$ is "**x** is the code of a Non-transferable proposition".(Also define $\boldsymbol{Tr}(\boldsymbol{x})$ as "**x** is the code of a Transferable proposition").

$\boldsymbol{NT(0)}$ means "For all codes **x** of the propositions $\boldsymbol{NT(x)}$, all are Non-transferable". This proposition can be constructed based on the structure of the channel as the following.

According to definition 2.3:

$$\begin{aligned}&\boldsymbol{NT}\left(\boldsymbol{F}_{\boldsymbol{P_n(m)}}\right) \text{ is equivalent to } \neg[\boldsymbol{Decode}(\boldsymbol{TS}\left(\boldsymbol{Code}(\boldsymbol{P_n}(\boldsymbol{m}))\right) \leftrightarrow \boldsymbol{P_n}(\boldsymbol{m})]\\ &\boldsymbol{Tr}\left(\boldsymbol{F}_{\boldsymbol{P_n(m)}}\right) \text{ is equivalent to } \boldsymbol{Decode}\left(\boldsymbol{TS}\left(\boldsymbol{Code}(\boldsymbol{P_n}(\boldsymbol{m}))\right)\right) \leftrightarrow \boldsymbol{P_n}(\boldsymbol{m})\end{aligned} \tag{3.1}$$

This means that the proposition $\boldsymbol{Tr}\left(\boldsymbol{F}_{\boldsymbol{P_n(m)}}\right)$ is true if and only if the received $\boldsymbol{P_n(m)}$ after coding and passing through the channel and decoding is equivalent to $\boldsymbol{P_n(m)}$, otherwise, $\boldsymbol{P_n(m)}$ is not transferrable on the channel or $\boldsymbol{NT}\left(\boldsymbol{F}_{\boldsymbol{P_n(m)}}\right)$ is true.

Suppose that transmitter has prepared $\boldsymbol{NT(0)}$ and has sent it toward the receiver. There are two statuses' in the receiver:

- $\boldsymbol{NT(0)}$ *is transferrable:* The decoded received proposition is equivalent to the $\boldsymbol{NT(0)}$, so the receiver interprets it as the non-transferability of all propositions including $\boldsymbol{NT(0)}$ itself and this is a contradiction.
- $\boldsymbol{NT(0)}$ *is not transferrable:* The decoded received proposition is not equivalent to the $\boldsymbol{NT(0)}$.

So we succeeded in constructing a proposition $\boldsymbol{NT(0)}$ that cannot be transferred in any case. ■

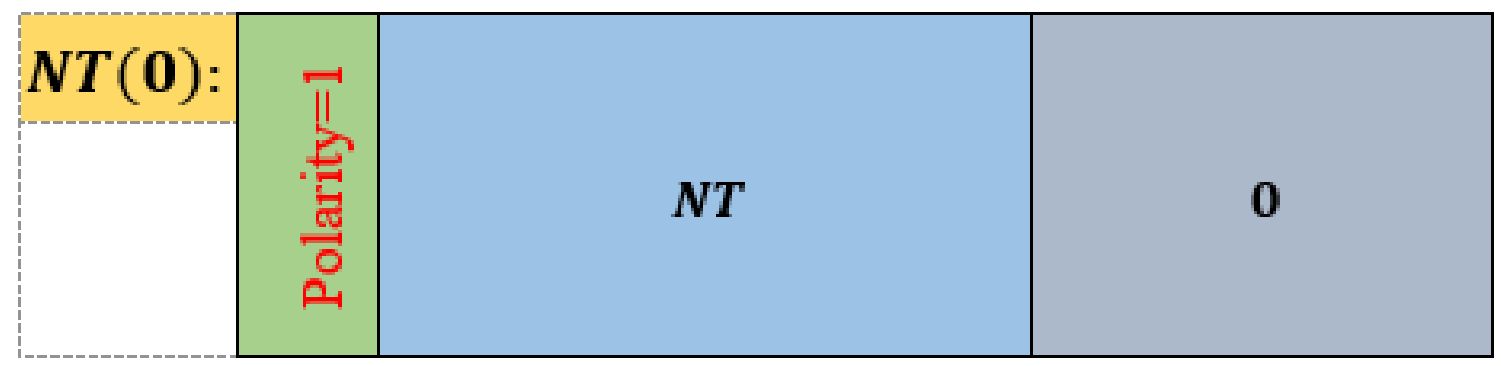

**Fig.5 The frame representation of NT(0) is prepared and sent over the channel**

***Proof B,*** *(Formal setup):*

Let's consider all different propositions or data frames and arrange all of them in a table as in Table 1 .

- $[\boldsymbol{1}, \boldsymbol{P_n}, \boldsymbol{m}]$ is equal to $\boldsymbol{P_n(m)}$**.** It means that object #**m** has property $\boldsymbol{P_n}$.
- $[\boldsymbol{0}, \boldsymbol{P_n}, \boldsymbol{m}]$ is equal to $\neg\boldsymbol{P_n(m)}$. It means that object #**m** has not property $\boldsymbol{P_n}$.

| | 1 | 2 | 3 | 4 | | m | |
|---|---|---|---|---|---|---|---|
| $\boldsymbol{P_1(m)}$: | $[\mathbf{1}, \boldsymbol{P_1}, 1]$ | $[\mathbf{1}, \boldsymbol{P_1}, 2]$ | $[\mathbf{1}, \boldsymbol{P_1}, 3]$ | $[\mathbf{1}, \boldsymbol{P_1}, 4]$ | ... | $[\mathbf{1}, \boldsymbol{P_1}, m]$ | ... |
| $\neg\boldsymbol{P_1(m)}$: | $[\mathbf{0}, \boldsymbol{P_1}, 1]$ | $[\mathbf{0}, \boldsymbol{P_1}, 2]$ | $[\mathbf{0}, \boldsymbol{P_1}, 3]$ | $[\mathbf{0}, \boldsymbol{P_1}, 4]$ | ... | $[\mathbf{0}, \boldsymbol{P_1}, m]$ | ... |
| $\boldsymbol{P_2(m)}$: | $[\mathbf{1}, \boldsymbol{P_2}, 1]$ | $[\mathbf{1}, \boldsymbol{P_2}, 2]$ | $[\mathbf{1}, \boldsymbol{P_2}, 3]$ | $[\mathbf{1}, \boldsymbol{P_2}, 4]$ | ... | $[\mathbf{1}, \boldsymbol{P_2}, m]$ | ... |
| $\neg\boldsymbol{P_2(m)}$: | $[\mathbf{0}, \boldsymbol{P_2}, 1]$ | $[\mathbf{0}, \boldsymbol{P_2}, 2]$ | $[\mathbf{0}, \boldsymbol{P_2}, 3]$ | $[\mathbf{0}, \boldsymbol{P_2}, 4]$ | ... | $[\mathbf{0}, \boldsymbol{P_2}, m]$ | ... |
| **:** | : | : | : | : | | : | |
| $\boldsymbol{P_n(m)}$: | $[\mathbf{1}, \boldsymbol{P_n}, 1]$ | $[\mathbf{1}, \boldsymbol{P_n}, 2]$ | $[\mathbf{1}, \boldsymbol{P_n}, 3]$ | $[\mathbf{1}, \boldsymbol{P_n}, 4]$ | ... | $[\mathbf{1}, \boldsymbol{P_n}, m]$ | ... |
| $\neg\boldsymbol{P_n(m)}$: | $[\mathbf{0}, \boldsymbol{P_n}, 1]$ | $[\mathbf{0}, \boldsymbol{P_n}, 2]$ | $[\mathbf{0}, \boldsymbol{P_n}, 3]$ | $[\mathbf{0}, \boldsymbol{P_n}, 4]$ | ... | $[\mathbf{0}, \boldsymbol{P_n}, m]$ | ... |
| : | : | : | : | : | : | : | : |
| | | | | | | | |
| $\boldsymbol{B(n)}$ | $[1, NT, [\mathbf{1}, \boldsymbol{P_1}, 1]]$ | $[1, NT, [\mathbf{1}, \boldsymbol{P_2}, 2]]$ | $[1, NT, [\mathbf{1}, \boldsymbol{P_3}, 3]]$ | $[1, NT, [\mathbf{1}, \boldsymbol{P_4}, 4]]$ | ... | $[1, NT, [\mathbf{1}, \boldsymbol{P_n}, n]]$ | |
| $\boldsymbol{B'(n)}$ | $[0, Tr, [\mathbf{0}, \boldsymbol{P_1}, 1]]$ | $[0, Tr, [\mathbf{0}, \boldsymbol{P_2}, 2]]$ | $[0, Tr, [\mathbf{0}, \boldsymbol{P_3}, 3]]$ | $[0, Tr, [\mathbf{0}, \boldsymbol{P_4}, 4]]$ | ... | $[0, Tr, [\mathbf{0}, \boldsymbol{P_n}, n]]$ | |

**Table 1. The enumeration of all propositions in the communication channel**

Since *Transferability* and also *Non-Transferability* are definable as in formula in 2.2, then we construct two special propositions or data frames named $\boldsymbol{B(n)}, \boldsymbol{B'(n)}$ based on them:

$$\begin{aligned} \boldsymbol{B(n)} &\equiv \boldsymbol{NT}\left(\boldsymbol{F_{P_n(n)}}\right) \equiv \left[1, NT, \boldsymbol{F}_{P_n(n)}\right] \equiv [1, NT, [\mathbf{1}, \boldsymbol{P_n}, n]] \\ \boldsymbol{B'(n)} &\equiv \neg\boldsymbol{Tr}\left(\boldsymbol{F_{\neg P_n(n)}}\right) \equiv [0, Tr, \boldsymbol{F}_{P_n(n)}] \equiv [0, Tr, [\mathbf{0}, \boldsymbol{P_n}, n]] \end{aligned} \tag{3.2}$$

For each **n** we construct the frame $[\mathbf{1}, \boldsymbol{P_n}, \boldsymbol{n}]$ (or $\boldsymbol{F_{P_n(n)}}$) and evaluate its non-transferability by the predicate $\boldsymbol{NT}(.)$ and set $\boldsymbol{B(n)}$ as the frame $[\mathbf{1}, \boldsymbol{NT}, [\mathbf{1}, \boldsymbol{P_n}, \boldsymbol{n}]]$. Also, we construct the frame $[\mathbf{0}, \boldsymbol{P_n}, \boldsymbol{n}]$ (or $\boldsymbol{F_{\neg P_n(n)}}$) and evaluate its Transferability by the predicate $\boldsymbol{Tr}(.)$ and set $\boldsymbol{B'(n)}$ as $[\mathbf{0}, \boldsymbol{Tr}, [\mathbf{0}, \boldsymbol{P_n}, \boldsymbol{n}]]$.

Based on the assumption that all propositions are in the list, then these two apparently new propositions must coincide with two rows in the table. There must be $\boldsymbol{k}$ **and** $\boldsymbol{k'}$ that:

$$\begin{aligned} \boldsymbol{B(n)} &\equiv \boldsymbol{P_k(n)} \\ \boldsymbol{B'(n)} &\equiv \neg\boldsymbol{P_{k'}(n)} \end{aligned} \tag{3.3}$$

So there are $\boldsymbol{k}$ **and** $\boldsymbol{k'}$ that for every **n**:

$$\begin{aligned} \boldsymbol{P_k(n)} &\equiv \boldsymbol{NT}\left(\boldsymbol{F_{P_n(n)}}\right) \quad \textbf{or} \quad [\mathbf{1}, \boldsymbol{P_k}, \boldsymbol{n}] \equiv \left[\mathbf{1}, \boldsymbol{NT}, [\mathbf{1}, \boldsymbol{P_n}, \boldsymbol{n}\right] \\ \neg\boldsymbol{P_{k'}(n)} &\equiv \neg\boldsymbol{Tr}\left(\boldsymbol{F_{\neg P_n(n)}}\right) \quad \textbf{or} \quad [\mathbf{0}, \boldsymbol{P_{k'}}, \boldsymbol{n}] \equiv \left[\mathbf{0}, \boldsymbol{Tr}, [\mathbf{0}, \boldsymbol{P_n}, \boldsymbol{n}\right] \end{aligned} \tag{3.4}$$

By selecting $\boldsymbol{n} = \boldsymbol{k}$ in the first formula and $\boldsymbol{n} = \boldsymbol{k'}$ in the second one:

$$
\begin{array}{lcl}
\boldsymbol{P_k(k) \equiv NT\left(F_{P_k(k)}\right)} & \textbf{or} & \boldsymbol{[1, P_k\, , k] \equiv \left[1, NT, [1, P_{k'}, k\right]} \\
\boldsymbol{\neg P_{k'}(k') \equiv \neg Tr\left(F_{\neg P_{k'}(k')}\right)} & \textbf{or} & \boldsymbol{[0, P_{k'}\, , k'] \equiv \left[0, Tr, [0, P_{k'}, k'\right]}
\end{array}
\qquad (3.5)
$$

The first formula means that $\boldsymbol{P_k(k)}$ as a true proposition implies that it is not transferable. Also, the second formula means that $\boldsymbol{\neg P_{k'}(k')}$ as a true proposition results in that $\boldsymbol{\neg P_{k'}(k')}$ is not transferable. So we succeeded in constructing propositions that are not transferable. The structure of this proof is similar to the famous diagonal method in set theory. ■

## 4. Non-Transferability and Tarski's Truth Theorem

Theorem 3.1 articulates within communication theory what Tarski's argument demonstrates in mathematical logic. According to the channel non-transferability theorem, there is no encoding-decoding mechanism capable of transmitting all propositions along with their truth value from the transmitter to the receiver. This corresponds to the non-existence of an arithmetical function or encoder-decoder that can represent all truths states of prepositions within the system, as stated by Tarski's truth theorem. the structure of the proof follows the fixed-point method, a renowned approach used in many logical theorems. Both theorems emphasized the impossibility of finding a method to 'express' or 'transfer' the truth values of all facts within the respective system. This equivalence is formally proven in the following theorem.

***Theorem 4.1****: The* non-transferability of an active communication channel and Tarski's truth undefinability theorem are equivalent.

***Proof:*** Suppose an arbitrary active channel $\boldsymbol{C} = (\boldsymbol{Encode}(.), \boldsymbol{TS}(.), \boldsymbol{Decode}(.))$ including an encoder and decoder to represent and send every proposition in the system. The channel is active, so there could be a one-to-one correspondence between encoder output $\boldsymbol{D}$ and decoder input $\boldsymbol{D'}$. The presence of suitable coding for representing each proposition and its truth value in the system is at the core of the proof. I use the method of "proof by contradiction":

a) If $\boldsymbol{C}$ has *Non-Transferability* and Tarski's truth theorem is not the case:
   Since it is supposed that Tarski's theorem is not valid, there is a 1-ary arithmetic predicate ***T(x)*** as the truth function. That is, the truth value of each proposition $\boldsymbol{P}$ can be evaluated by the truth value of ***T(n)*** in which **n** is Gödel's number or code of $\boldsymbol{P}$.

$$\boldsymbol{T}(\boldsymbol{n}) \leftrightarrow \boldsymbol{P} \quad \text{and} \quad \boldsymbol{n} = \boldsymbol{Encode}(\boldsymbol{P})$$
$$\text{For each proposition } \mathbf{P}\text{:} \quad \boldsymbol{T}(\boldsymbol{Encode}(\boldsymbol{P})) \leftrightarrow \boldsymbol{P} \tag{4.1}$$

Let's assume that we are going to send an arbitrary proposition $\boldsymbol{P}$ toward the receiver. To do this, the same coding mechanism as is used to build **n** from $\boldsymbol{P}$ in 4.1 is selected. After inputting **n** to the channel, it delivers a corresponding unique $\boldsymbol{n'}$ to the input of the decoder based on the one-to-one function $\boldsymbol{TS}(.)$ (This derives from the activeness of the channel).

$$\boldsymbol{n'} = \boldsymbol{TS}(\boldsymbol{n}) \quad \boldsymbol{TS} \text{ is a one-to-one function}$$
$$\text{So} \quad \boldsymbol{n'} = \boldsymbol{TS}(\boldsymbol{Encode}(\boldsymbol{P})) \tag{4.2}$$

Since the function ***T(x)*** is definable, the decoder function is definable too:

$$\boldsymbol{Decode}(.) \equiv \boldsymbol{T}(\boldsymbol{TS}^{-1}(.)) \tag{4.3}$$

The final output of the system is:

$$\begin{aligned} \boldsymbol{P'} &\leftrightarrow \boldsymbol{Decode}(\boldsymbol{n'}) \\ &\leftrightarrow \boldsymbol{T}\left(\boldsymbol{TS}^{-1}(\boldsymbol{n'})\right) \\ &\leftrightarrow \boldsymbol{T}\left(\boldsymbol{TS}^{-1}(\boldsymbol{TS}(\boldsymbol{Encode}(\boldsymbol{P})))\right) \\ &\leftrightarrow \boldsymbol{T}(\boldsymbol{Encode}(\boldsymbol{P})) \\ &\leftrightarrow \boldsymbol{P} \end{aligned} \tag{4.4}$$

The proposition $\boldsymbol{P'}$ as the output of the channel is equivalent to $\boldsymbol{P}$. So the truth value of $\boldsymbol{P}$ after transferring on the channel can be determined based on the arithmetical predicate ***T(x)***. In another word, that channel has transferability for all propositions. → Contradiction.

b) If Tarski's truth theorem is the case and ***C*** has *Transferability*:
In this case, every proposition $\boldsymbol{P}$ in the system can be encoded in a frame, through arithmetic operations, and can be decoded unambiguously in the receiver as $\boldsymbol{P'}$ whose truth value is the same as $\boldsymbol{P}$**:**

$$\boldsymbol{P'} \leftrightarrow \boldsymbol{Decode}\left(\boldsymbol{TS}\left(\boldsymbol{Encode}(\boldsymbol{P})\right)\right) \tag{4.5}$$

So by this framing and coding-decoding mechanism, we can represent every proposition of the system containing its truth value by a number **n**:

$$\boldsymbol{n} = \boldsymbol{Encode}(\boldsymbol{P})$$
$$\boldsymbol{n} = \boldsymbol{F_P} \qquad \textbf{n} \text{ is like a binary representation of the Frame } \boldsymbol{F_P} \qquad (4.6)$$

Let's define predicate $\boldsymbol{T}(\boldsymbol{n})$ as below:

$$\boldsymbol{T}(\boldsymbol{n}) \equiv \boldsymbol{Decode}\big(\boldsymbol{TS}(\boldsymbol{n})\big) \qquad (4.7)$$

Then:

$$\boldsymbol{T}(\boldsymbol{n}) \equiv \boldsymbol{Decode}\big(\boldsymbol{TS}(\boldsymbol{n})\big)$$
$$\boldsymbol{T}(\boldsymbol{n}) \equiv \boldsymbol{Decode}\Big(\boldsymbol{TS}\big(\boldsymbol{Encode}(\boldsymbol{P})\big)\Big) \qquad (4.8)$$
$$\boldsymbol{T}(\boldsymbol{n}) \equiv \boldsymbol{P}'$$

But since $\boldsymbol{P}' \leftrightarrow \boldsymbol{P}$ then:

$$\boldsymbol{P} \leftrightarrow \boldsymbol{T}(\boldsymbol{n}) \quad \& \quad \boldsymbol{n} = \boldsymbol{Encode}(\boldsymbol{P}) \qquad (4.9)$$

So we have an arithmetic function $\boldsymbol{T}(\boldsymbol{n})$ that can give the truth value of each proposition, and this is not what Tarski's theorem is saying. → Contradiction.

These two parts of the proof collectively conclude that non-transferability and Tarski's undefinability of truth are equivalent. In essence, the existence of a channel capable of transmitting every proposition along with its truth value is equivalent to having an arithmetic function that represents a complete truth table for all propositions. ■

Now, let us examine the more fundamental relationships between a communication channel and Tarski's definition of truth.

Based on *Fig.2*, it is going to inform the receiver about the set of facts $\boldsymbol{S}$. Also, $\boldsymbol{S}'$ is the set of facts constructed by the receiver based on massage $\boldsymbol{M}'$. If $\boldsymbol{S}'$=$\boldsymbol{S}$ then the communication channel and related systems are errorless. The receiver, in fact, is going to check the truth value of the elements of $\boldsymbol{S}$ and it does it through checking the received $\boldsymbol{S'}$:

$$\boldsymbol{S} \textit{ is true if and only if } \boldsymbol{S'}. \tag{4.10}$$

(Here "**S** is true" means "all elements of **S** are true")

This sentence is similar to the famous T-scheme definition of truth by Tarski:

$$\boldsymbol{P} \textit{ is true if and only if } \boldsymbol{P} \textit{ is the case.} \tag{4.11}$$

The communication channel, in fact, behaves like an interpretation system. The receiver is like a cognitive agent that is going to have all truth about all propositions and does it through the information that gets from the transmitter using the channel. So, the Tarski's definition of truth as in **(4.11)** can be expressed by the channel model of communication system as in **(4.10)**.

Next, an immediate consequence of non-transferability in communication channels is demonstrated, providing deeper understanding into the relationship between non-transferability in communication channels and Tarski's theorem.

The Liar Paradox, a well-known self-referential scenario is used for this demonstration. First a Liar Paradox situation is modeled within the system, and then it is examined whether the communication channel can manage it. The below theorem represents a non-transferable situation within a channel that is equivalent to the Liar Paradox.

***Theorem** 4.2:* In one use of a communication channel, the error situation of the channel is not transferable.

***Proof:*** Consider the situation in *Figure 6* , in which I am going to transmit channel error in one use of the channel:

$$\begin{aligned} &\boldsymbol{S_e} = \{\boldsymbol{The\ channel\ has\ error}\} \\ &\boldsymbol{S_e} = \{\boldsymbol{D} = \boldsymbol{Encode}(\boldsymbol{M}) \ \wedge \ \boldsymbol{TS}(\boldsymbol{D}) = \boldsymbol{D'} \ \wedge \ \boldsymbol{M'} = \boldsymbol{Decode}(\boldsymbol{D'}) \wedge \boldsymbol{M'} \neq \boldsymbol{M}\} \end{aligned} \tag{4.12}$$

- "$\boldsymbol{D} = \boldsymbol{Encode}(\boldsymbol{M})$" means that $\boldsymbol{D}$ is the encoded form of the message $\boldsymbol{M}$.
- "$\boldsymbol{M'} = \boldsymbol{Decode}(\boldsymbol{D'})$" is the inverse function of $\boldsymbol{Encode}(.)$ Function.
- "$\boldsymbol{TS}(\boldsymbol{D}) = \boldsymbol{D'}$" means that $\boldsymbol{D'}$ is the received string of the **T**ransmission **S**ystem with input $\boldsymbol{D}$.
- "$\boldsymbol{M'} \neq \boldsymbol{M}$" means that the channel has error and the message $\boldsymbol{M}$ is not transferable on it.

This situation is used when the transmitter wants to inform the receiver that the channel has error in transmitted data.

To model this situation, let us make a simple representation of it. Consider a simple proposition named ***Err(x)*** and it means that channel has error in sending string **x**:

- $\boldsymbol{Err(a)}$ ***means that channel has error in sending a special string*** "$\boldsymbol{a}$".
- $\boldsymbol{Err(0)}$ ***means that channel has error in sending each string.***[4]

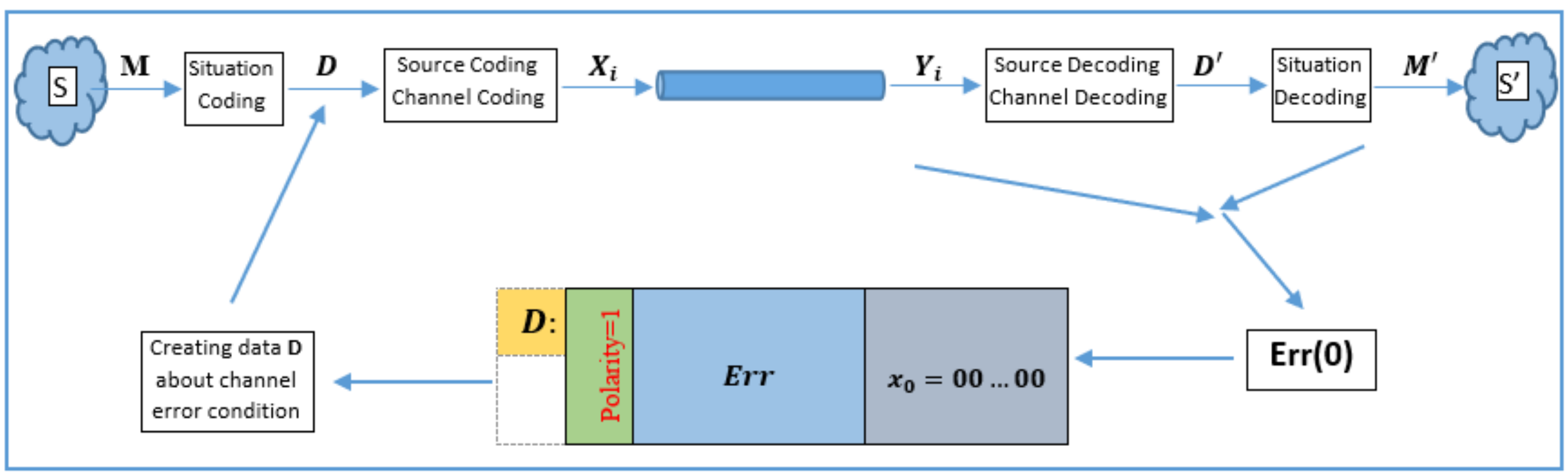

**Fig.6 Modeling a channel to communicate its error situation**

Now suppose that the transmitter encodes the proposition $\boldsymbol{Err(0)}$ and sends it toward the receiver. So the string is decodable and the receiver extracts the message $\boldsymbol{Err(0)}$ with polarity 1. So the received frame is saying that the received message $\boldsymbol{Err(0)}$ is true.

➔ $\boldsymbol{Err(0)}$ is true → Channel had error for all strings → $\boldsymbol{Err(0)}$ itself is not true

So, the status of the channel is not known or decidable to the receiver at all. That's the way, we succeeded in finding a situation that is not transferable at all. ■

As revealed in this proof, a communication system cannot transmit some situations that are referring to its error status. The theorem that is proved is the communication counterpart of the liar paradox reasoning.

---

[4] It is common in telecommunication systems to model the error condition of the channel by a special code. For example, in SDH transmission systems, this condition is shown by a frame with all 1 in its payload.

## 5. Conclusion

As demonstrated in this article, the existence of non-transferable codes in the field of communication theory is equivalent to the undefinability of truth as outlined by Tarski's theorem in mathematical logic. Similarly, the existence of non-computable functions in computer science is analogous to the result of Gödel's first incompleteness theorem in mathematical logic. These two equivalences are illustrated in *Figure 7* .

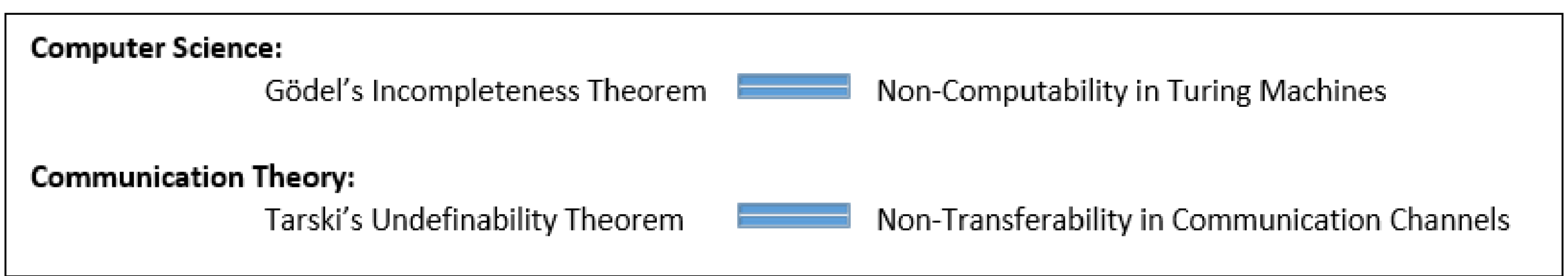

**Fig.7 Famous logical theorems and their practical counterparts**

It is well established that Gödel's incompleteness theorem reveals inherent limitations within any computing system. Tarski's undefinability of truth theorem, regarded as the most significant result in mathematical logic following Gödel's work, imposes another fundamental limitation, as demonstrated by the findings of this article. However, while Gödel's theorem addresses the computational and reasoning capabilities of systems, Tarski's theorem highlights limitations within communications systems. Just as Gödel's incompleteness theorem signifies the boundaries of computational systems, Tarski's undefinability of truths reveals the intrinsic constraints in communication channels.

Moreover, Gödel's theorem points to the fundamental boundaries of human reasoning when handling formal systems. Similarly, Tarski's theorem underscores the limitations of human language in conveying and comprehending truth about the state of affairs in the world. Human cognitive abilities, supported by numerous mental mechanisms, are limited by the constraints outlined in the seminal works of Gödel's and Tarski, particularly in reasoning and language.

From a system theory perspective, the existence of non-computable functions and non-transferable situations imposes analogous restrictions on the computational and communicational capacities of any system.

This article not only elucidates the fundamental limitations inherent in communication systems but also offers a fresh perspective on Tarski's undefinability theorem according to Theorems 2 and 3 in this article, both the issue of non-transferability and undefinability of truth points to the same perceptual and linguistic constraints within the cognitive systems.